\documentstyle [12pt] {article}

\begin{document}
\begin{flushright}
December 1995\\
UDHEP-12-95\\
IFP-603-UNC\\
\end{flushright}
\begin{center}
{\large {\bf Chiral Symmetry Breaking by a Magnetic Field
in Weak-coupling QED}}\\
\vspace*{0.2 cm}
\vspace*{0.2 cm}
{\bf C. N. Leung$^{1,2,*}$ , Y. J. Ng$^{2,3,4,\dagger}$, and
A. W. Ackley$^2$}\\
\end{center}
\vspace*{0.1 cm}
\noindent{$^1$Department of Physics and Astronomy, University of Delaware,\\
  Newark, DE  19716 \\
$^2$Institute of Field Physics, Department of Physics and Astronomy, \\
  University of North Carolina, Chapel Hill, NC  27599 \\
$^3$Center for Theoretical Physics, M.I.T., Cambridge, MA  02139 \\
$^4$School of Natural Sciences, Institute for Advanced Study, \\
Princeton, NJ  08540 \\
\\
$^*$email: cnleung@chopin.udel.edu \\
$^{\dagger}$email: Ng@Physics.UNC.edu}
\vspace*{0.3 cm}
\begin{center}
{\bf ABSTRACT} \\
\end{center}

Using the nonperturbative Schwinger-Dyson equation, we show
that chiral symmetry in weak-coupling massless QED is dynamically
broken by a constant but arbitrarily strong external magnetic field.

\vspace*{0.3 cm}
\noindent
{PACS numbers: 11.30.Rd, 11.30.Qc, 12.20.Ds}
\newpage

Chiral symmetry plays an important role in elementary particle
and nuclear physics.  In this Letter we examine its breaking in the
theory of quantum electrodynamics.  It has been known\cite{Mir}
for some time that QED may have a nonperturbative strong-coupling
phase, characterized by spontaneous chiral symmetry breaking, in
addition to the familiar weak-coupling phase.  The existence of this
new phase was exploited in a novel interpretation\cite{newphase} of
the multiple correlated and narrow-peak structures in electron and
positron spectra\cite{GSI} observed at GSI several years ago.
According to this scenario, the $e^+e^-$ peaks are due to the decay
of a bound $e^+e^-$ system formed in the new QED phase, which is
induced by the strong and rapidly varying electromagnetic fields
present in the neighborhood of the colliding heavy ions.  While the
experimental situation with regard to these anomalous $e^+e^-$
events is unclear, especially after similar experiments at Argonne
have yielded negative results\cite{Argonne}, it is still of great interest
to investigate whether background fields can be physically used to
induce chiral symmetry breaking.  Now the question is: what kind
of background fields can potentially induce chiral symmetry breaking
in gauge theories?\cite{Ng}  We recall that in a magnetic monopole
field a gauge field breaks chiral symmetry\cite{BCT} and that in the
Nambu-Jona-Lasinio model a magnetic field drives the critical
transition point towards weaker coupling\cite{KL}.  Thus, magnetic
fields are obvious candidates.  We will take advantage of the
strong-field techniques introduced by Schwinger and others to consider
a constant magnetic field of arbitrary strength.  In order to put our
problem in as general a setting as possible, we will use the
nonperturbative Schwinger-Dyson equation approach.  We will
then compare our result with that obtained recently by Gusynin,
Miransky, and Shovkovy\cite{GMS} whose method is very different
from ours.  A comparison of the two approaches will sharpen our
understanding of the underlying physics and the kind of
approximations involved.

The motion of a massless fermion of charge $e$ in an external
electromagnetic field is described by the Green's function that
satisfies the modified Dirac equation proposed by Schwinger:
\begin{equation}
\gamma \cdot \Pi(x) G_A(x,y) + \int d^4x' M(x,x') G_A(x',y) =
\delta^{(4)}(x-y),
\label{Greeneq}
\end{equation}
where $\Pi_\mu(x) = - i \partial_\mu - e A_\mu(x)$, and $M(x,x')$
is the mass operator $M$ in the coordinate representation.  For a
constant magnetic field of strength $H$, we may take $A_2 = Hx_1$
to be the only nonzero component of $A_\mu$.  In the following we
will use the method due to Ritus\cite{Ritus}, which is based on the
use of the eigenfunctions of the mass operator and the diagonalization
of the latter.  As shown by Ritus, $M$ is diagonal in the
representation of the eigenfunctions $E_p(x)$ of the operator
$(\gamma \cdot \Pi)^2$:
\begin{equation}
- (\gamma \cdot \Pi)^2  E_p(x) = p^2 E_p(x).
\label{eigeneq}
\end{equation}
The advantage of using this representation is obvious: $M$ can now
be put in terms of its eigenvalues, so the problems arising from its
dependence on the operator $\Pi$ can be avoided.  In the chiral
representation in which $\sigma_3$ and $\gamma_5$ are diagonal
with eigenvalues $\sigma = \pm 1$ and $\chi = \pm 1$, respectively,
the eigenfunctions $E_{p\sigma\chi}(x)$ take the form
\begin{equation}
E_{p\sigma\chi}(x) = N {\rm e}^{i (p_0x^0 + p_2x^2 + p_3x^3)} D_n(\rho)
\omega_{\sigma\chi} \equiv \tilde{E}_{p\sigma\chi} \omega_{\sigma\chi},
\label{eigenfcn}
\end{equation}
where $D_n(\rho)$ are the parabolic cylinder functions\cite{math}
with indices
\begin{equation}
n = n(k,\sigma) \equiv k + \frac{e H \sigma}{2 |e H|} - \frac{1}{2},
{}~~~~k = 0, 1, 2, ...,
\label{index}
\end{equation}
and argument $\rho = \sqrt{2 |e H|} (x_1 - \frac{p_2}{e H})$.
Note that $n = 0,~1,~2,~...~$.
The normalization factor is $N = (4 \pi |eH|)^{1/4}/\sqrt{n!}$;
$p$ stands for the set $(p_0, p_2, p_3, k)$; and
$\omega_{\sigma\chi}$ are the bispinors of $\sigma_3$ and
$\gamma_5$.

Following Ritus, we form the orthonormal and complete\cite{ortho}
eigenfunction-matrices $E_p = {\rm diag}(\tilde{E}_{p11},~
\tilde{E}_{p-11},~\tilde{E}_{p1-1},~\tilde{E}_{p-1-1})$.  They
satisfy
\begin{equation}
\gamma \cdot \Pi~E_p(x) = E_p(x)~\gamma \cdot \bar{p}
\end{equation}
and
\begin{equation}
M(x,x') E_p(x') = E_p(x) \delta^{(4)}(x-x') \tilde{\Sigma}_A(\bar{p}),
\label{masseigeneq}
\end{equation}
where $\tilde{\Sigma}_A(\bar{p})$ represents the eigenvalues of
the mass operator, and $\bar{p}_0 = p_0,~\bar{p}_1 = 0,~\bar{p}_2
= - {\rm sgn}(eH) \sqrt{2|eH|k},~\bar{p}_3 = p_3$.  These
properties of the $E_p(x)$ allow us to express the Green's function
and the mass operator in the $E_p$-representation as
$(\bar{E}_p \equiv \gamma^0 E_p^\dagger \gamma^0)$
\begin{equation}
G_A(x,y) = \Sigma \!\!\!\!\!\! \int \frac{d^4p}{(2 \pi)^4} E_p(x) \frac{1}
{\gamma \cdot \bar{p} + \tilde{\Sigma}_A(\bar{p})} \bar{E}_p(y),
{}~~~\Sigma \!\!\!\!\!\! \int d^4p \equiv \sum_{k} \int dp_0 dp_2 dp_3,
\label{Greenfcn}
\end{equation}
and
\begin{equation}
M(p,p') = \int d^4x d^4x' \bar{E}_p(x) M(x,x') E_{p'}(x')
= \tilde{\Sigma}_A(\bar{p}) (2 \pi)^4 \hat{\delta}^{(4)}(p-p'),
\label{pmassop}
\end{equation}
respectively, where $\hat{\delta}^{(4)}(p-p') \equiv \delta_{kk'}
\delta(p_0 - p'_0) \delta(p_2 - p'_2) \delta(p_3 - p'_3)$.

We work in the ladder approximation in which
\begin{equation}
M(x,x') = i e^2 \gamma^\mu G_A(x,x') \gamma^\nu D_{\mu\nu}(x-x'),
\label{massop}
\end{equation}
where $D_{\mu\nu}(x-x')$ is the bare photon propagator,
\begin{equation}
D_{\mu\nu}(x-x') = \int \frac{d^4q}{(2 \pi)^4} \frac{{\rm e}^{i q\cdot(x-x')}}
{q^2 - i\epsilon} \left(g_{\mu\nu} -(1 - \xi) \frac{q_\mu q_\nu}{q^2}\right).
\label{photon}
\end{equation}
The Schwinger-Dyson (SD) equation then takes the form
\begin{eqnarray}
\tilde{\Sigma}_A(\bar{p}) (2 \pi)^4 \hat{\delta}(p-p')
&=&
i e^2 \int d^4x d^4x' \Sigma \!\!\!\!\!\! \int \frac{d^4p"}{(2 \pi)^4}
\bar{E}_p(x) \gamma^\mu E_{p"}(x) \nonumber \\
& &
\!\!\!\!\!\!\!\!\!\!\!\!\!\!\!\!\! \times \frac{1}{\gamma \cdot \bar{p"} +
\tilde{\Sigma}_A(\bar{p"})} \bar{E}_{p"}(x') \gamma^\nu E_{p'}(x')
D_{\mu\nu}(x-x').
\label{SDfull}
\end{eqnarray}

After integrations over $x,~x',~p"_{\!\!\!\!_0},~p"_{\!\!\!\!_2},$
and $p"_{\!\!\!\!_3}$, the SD equation is simplified to read
($r \equiv \sqrt{(q_1^2 + q_2^2)/(2|eH|)},~\varphi \equiv
\tan^{-1}(- q_2/q_1)$)
\begin{eqnarray}
\tilde{\Sigma}_A(\bar{p}) \delta_{kk'}
&=&
i e^2 \sum_{k"} \int \frac{d^4q}{(2 \pi)^4}
\frac{1}{\sqrt{n!n'!n"!\tilde{n}"!}}
{\rm e}^{- r^2} {\rm e}^{i~{\rm sgn}(eH)(n'-n+n"-\tilde{n}")\varphi}
\nonumber \\
& &
\times \frac{1}{q^2} \left(g_{\mu\nu} - (1 - \xi) \frac{q_\mu q_\nu}
{q^2}\right) \gamma^0 \Delta \gamma^0 \gamma^\mu \Delta" \nonumber \\
& &
\times \frac{1}{\gamma \cdot \bar{p"} + \tilde{\Sigma}_A(\bar{p"})}
\gamma^0 \tilde{\Delta}" \gamma^0 \gamma^\nu \Delta' J_{nn"} (r)
J_{\tilde{n}"n'}(r),
\label{SDsimp}
\end{eqnarray}
where summing over $\sigma,~\sigma',~\sigma"$, and
$\tilde{\sigma}"$ on the right hand side is understood, and
\begin{equation}
J_{nn'}(r) \equiv \sum_{m=0}^{{\rm min} (n,n')} \frac{n! n'!}{m! (n-m)!
(n'-m)!} [i~{\rm sgn}(eH)r]^{n+n'-2m}.
\label{defJ}
\end{equation}
We have also used the following notations\cite{chi} in
Eq.(\ref{SDsimp}):
$\bar{p"}_{\!\!\!\!_0} = p_0 - q_0$, $\bar{p"}_{\!\!\!\!_1} = 0$,
$\bar{p"}_{\!\!\!\!_2} = -~{\rm sgn}(eH) \sqrt{2|eH|k"}$,
$\bar{p"}_{\!\!\!\!_3} = p_3 - q_3$, $\Delta = \Delta(\sigma)
= {\rm diag} (\delta_{\sigma 1}, \delta_{\sigma -1},
\delta_{\sigma 1}, \delta_{\sigma -1})$, $\Delta' = \Delta(\sigma')$,
 ... etc., $n' = n(k',\sigma')$, $n" = n(k",\sigma")$, and $\tilde{n}" =
n(k",\tilde{\sigma}")$.

Eq.(\ref{SDsimp}) may be solved by following the standard
procedure\cite{HN} of writing $\tilde{\Sigma}_A(\bar{p}) =
\beta \gamma \cdot \bar{p} + \Sigma_A(\bar{p})$, where
$\Sigma_A(\bar{p})$ corresponds to the dynamically
generated fermion mass.  We will assume that
$\Sigma_A(\bar{p})$ is proportional to the unit matrix
(it will be seen later from the solution that this is a
self-consistent assumption).  Eq.(\ref{SDsimp}) then leads to two
coupled equations for $\beta$ and $\Sigma_A$:
\begin{eqnarray}
\left( \begin{array}{c} \Sigma_A(\bar{p}) \\
\beta \gamma \cdot \bar{p} \end{array} \right) \delta_{kk'}
& = &
i e^2 \sum_{k"} \int \frac{d^4q}{(2 \pi)^4} \frac{1}
{\sqrt{n!n'!n"!\tilde{n}"!}} {\rm e}^{- r^2}
{\rm e}^{i~{\rm sgn}(eH)(n'-n+n"-\tilde{n}")\varphi} \nonumber \\
& &
\times J_{nn"} (r) J_{\tilde{n}"n'}(r) \frac{1}
{(1+\beta)^2 \bar{p"}^2 + \Sigma_A^2(\bar{p"})} \nonumber \\
& &
\times \frac{1}{q^2} \left( \begin{array}{c} \Sigma_A(\bar{p"})
(G_1 - \frac{1-\xi}{q^2} Q_1) \\
- (1+\beta) (G_2 - \frac{1-\xi}{q^2} Q_2) \end{array} \right)
\label{SDcouple}
\end{eqnarray}
where
$G_1 = \Delta \gamma^\mu \Delta" \tilde{\Delta}"
\gamma_\mu \Delta' = - 2 (\delta_{\sigma"1}\delta_{\tilde{\sigma}"1}
+ \delta_{\sigma"-1}\delta_{\tilde{\sigma}"-1}) {\rm diag}
(\delta_{\sigma 1}\delta_{\sigma' 1}, \delta_{\sigma -1}
\delta_{\sigma' -1}$, $\delta_{\sigma 1}\delta_{\sigma' 1},
\delta_{\sigma -1}\delta_{\sigma' -1})$, $Q_1 = \Delta (\gamma
\cdot q) \Delta" \tilde{\Delta}" (\gamma \cdot q) \Delta'$,
$G_2 = \Delta \gamma^\mu \Delta" (\gamma \cdot \bar{p"})
\tilde{\Delta}" \gamma_\mu \Delta'$, and $Q_2 = \Delta
(\gamma \cdot q) \Delta" (\gamma \cdot \bar{p"})
\tilde{\Delta}" (\gamma \cdot q) \Delta'$.

We seek solutions with $\beta = 0$.  We will show later
that such a solution is consistent only with the Feynman
gauge ($\xi = 1$).  In this case the two SD equations
decouple and only $G_1$ is relevant for determining the
dynamical fermion mass.  The spin structure of $G_1$
implies that necessarily $\sigma" = \tilde{\sigma}"$,
which, in turn, implies that necessarily $n" = \tilde{n}"$.
It is convenient to make a change of integration variables
from $(q_1, q_2)$ to the "polar coordinates" $(r, \varphi)$.
The integration over $\varphi$ yields
\begin{equation}
\int_{0}^{2 \pi} d\varphi {\rm e}^{i~{\rm sgn}(eH) (n'-n) \varphi}
= 2 \pi \delta_{nn'}
\label{phiint}
\end{equation}
We note that the spin structure of $G_1$ also implies that $\sigma
= \sigma'$, which, together with the $\delta_{nn'}$ from
Eq.(\ref{phiint}), matches the $\delta_{kk'}$ on the left hand
side of Eq.(\ref{SDcouple}).

Due to the factor e$^{- r^2}$ in the integrand in Eq.(\ref{SDcouple}),
contributions from large values of $r$ are suppressed.  Let us therefore,
as an approximation (we will find out later what physical condition
validates this approximation), keep only the smallest power of $r$ in
$J_{nn"}(r)$, i.e.,
\begin{equation}
J_{nn"}(r) \rightarrow \frac{[{\rm max}(n, n")]!}{|n-n"|!}
(i~{\rm sgn}(eH)r)^{|n-n"|}.
\end{equation}
Since the leading contributions come from the term corresponding to
$n" = n$, we need only keep the term given by $k" = n + \frac{1}{2}
- \frac{\sigma"}{2} {\rm sgn}(eH)$ in the summation over $k"$.
As a result, we can replace $J_{nn"}$ by $n!$.  The SD equation
(Eq.(\ref{SDcouple})), thereby vastly simplified, becomes
\begin{equation}
\Sigma_A(\bar{p}) \simeq \frac{i e^2}{(2 \pi)^3} |eH| \int dq_0 dq_3
\int_0^\infty dr^2 {\rm e}^{- r^2} \frac{G_1}{q^2} \frac
{\Sigma_A(\bar{p"})}{\bar{p"}^2 + \Sigma_A(\bar{p"})}
\label{fermass}
\end{equation}
where $q^2 = - q_0^2 + q_3^2 + 2 |eH| r^2$ and $\bar{p"}^2 =
- (p_0 - q_0)^2 + (p_3 - q_3)^2 + 2 |eH| k"$.

Let us make a Wick rotation to Euclidean space: $p_0 \rightarrow i p_4$,
$q_0 \rightarrow i q_4$.  Consider the case with $p = 0$, i.e., $p_0 = p_3
= k = 0$.  Notice that $k = 0$ means that, for positive (negative) $eH$,
$\sigma = 1 (-1)$ and $n = 0$, the last of which implies that $k" = 0$
and $\sigma" = 1 (-1)$ for the respective sign of $eH$.  We also note
that for either sign of $eH$, the matrix $G_1$ can be effectively
replaced by $-2 \times {\bf 1}$.  We will assume that the dominant
contributions to the integral in Eq.(\ref{fermass}) come from the
infrared region of small $q_3$ and $q_4$ (this assumption will
be seen to be self-consistent).  Thus, it is reasonable to replace
$\Sigma_A(\bar{p"})$ in the integrand by $\Sigma_A(0) = m \times
{\bf 1}$.  Eq.(\ref{fermass}) then becomes
\begin{equation}
m \simeq \frac{\alpha}{\pi^2} \int dq_3 dq_4 \int_0^\infty dr^2
{\rm e}^{- r^2} \frac{1}{2 r^2 + L^2 (q_3^2 + q_4^2)}~
\frac{m}{m^2 + (q_3^2 +q_4^2)}
\label{dynmass}
\end{equation}
where $\alpha = e^2/4 \pi$ is the fine structure constant and $L =
1/\sqrt{|eH|}$ is the magnetic length.  The integrations over $q_3$
and $q_4$ give
\begin{equation}
1 \simeq \frac{\alpha}{\pi} \int_0^\infty dr^2 \frac{{\rm e}^{- r^2}
\ln(2 r^2/m^2 L^2)}{2 r^2 - m^2 L^2}
\label{masscond}
\end{equation}
which yields the nonzero dynamical mass as
\begin{equation}
m \simeq a~\sqrt{|eH|}~{\rm e}^{- b\sqrt{\frac{\pi}{\alpha} - c}},
\label{result}
\end{equation}
where $a$, $b$, and $c$ are constants of order 1.

Eq.(\ref{result}) clearly demonstrates the nonperturbative nature of
the result.  It also shows that our approximations break down when
$\alpha > O(1)$.  As a further check on the consistency of
our assumptions, we note that, according to Eq.(\ref{dynmass}),
the dominant contributions to the integrals come from the region
$2 r^2 \sim m^2 L^2 \sim L^2 (q_3^2 +q_4^2)$.  Our earlier
assumption that effectively $r \ll 1$ is now translated to the
physical assumption that $m L \ll 1$, which requires that
$\alpha \ll O(1)$; in other words, the dynamical chiral
symmetry breaking solution we have found applies to the
weak-coupling regime of QED!  Now it is also evident that indeed
the infrared region of $q_3$ and $q_4$ gives the dominant
contributions to the integrals.

It remains for us to show that $\beta = 0$ solves Eq. (\ref{SDcouple})
only if $\xi = 1$.  The main point to note is that, consistent with the
$k" = 0$ approximation made above, we can approximate the $\gamma
\cdot \bar{p"}$ in $G_2$ by $-(\gamma^0 q_0 + \gamma^3 q_3)$ (recall
also that we are considering the case of $p = 0$).  But then the piece of
the integrand involving $G_2$ is odd in $q_0$ as well as in $q_3$,
and hence vanishes upon integration.  It follows that the solution
$\beta = 0$ requires $\xi = 1$, i.e., the Feynman gauge.  As a result, the
$Q_1$-piece on the right hand side of Eq. (\ref{SDcouple}) does not
contribute.  (That is fortunate because $Q_1$ is actually not proportional
to the unit matrix; its presence in the SD equation would have spoiled
the assumption that $\Sigma_A$ is the dynamical mass multiplied by
the unit matrix.)

In summary, we have found a solution to the Schwinger-Dyson equation
in the presence of an arbitrarily strong constant magnetic field, which
indicates that, even at weak gauge coupling, an external magnetic field
can trigger the dynamical breaking of chiral symmetry in QED, with the
dynamical mass of the fermion given by Eq.(\ref{result}).  Our general
conclusion
agrees with a recent finding by Gusynin {\it et al.}\cite{GMS}, whose
approach is very different from ours.  It would be interesting to examine
if there are additional solutions of chiral symmetry breaking due to an
external magnetic field.  A parallel calculation for the case of a constant
background electric field\cite{DW} or other background field
configurations may also shed light on the dynamics of chiral symmetry
breaking in gauge theories.  The formalism proposed here will be most
suitable for these studies.

\vspace*{1.0cm}

\noindent{{\bf Acknowledgement.}  This work was supported in part
by the U.S. Department of Energy under Grant No. DE-FG02-84ER40163
and DE-FG05-85ER-40219 Task A, and by the Bahnson Fund and the Reynolds
Fund of UNC.
The work reported here was done some time ago when C.N.L. was visiting
UNC and when Y.J.N. was visiting M.I.T. and the Institute for Advanced
Study.  They thank the respective faculties for their hospitality.  C.N.L.
also thanks Tim Ziman, and Y.J.N. thanks S. Adler,
W. Chen, K. Johnson, K. Milton, and G. Semenoff for
useful conversations.  Y.J. Ng also thanks the late J.S. Schwinger for a
useful discussion in March 1988, to whose memory this paper is dedicated.}

\raggedbottom


\newpage


\begin{thebibliography}{99}


\bibitem{Mir}
T. Maskawa and H. Nakajima, Prog. Theor. Phys. {\bf 52}, 1326 (1974);
{\it ibid}., {\bf 54}, 860 (1975); R. Fukuda and T. Kugo, Nucl. Phys.
{\bf B117}, 250 (1976); V.A. Miransky, Il. Nuovo Cim. {\bf 90A}, 149
(1985); W.A. Bardeen, C.N. Leung, and S.T. Love, Nucl. Phys. {\bf B273},
649 (1986); {\it ibid}., {\bf B323}, 493 (1989); K. Yamawaki, M. Bando,
and K. Matumoto, Phys. Rev. Lett. {\bf 56}, 1335 (1986); J.B. Kogut,
E. Dagotto, and A. Kocic, Phys. Rev. Lett. {\bf 60}, 772 (1988).

\bibitem{newphase}
Y.J. Ng and Y. Kikuchi, Phys. Rev. D {\bf 36}, 2880 (1987); D.G. Caldi
and A. Chodos, Phys. Rev. D {\bf 36}, 2876 (1987); L.S. Celenza, V.K. Mishra,
C.M. Shakin, and K.F. Liu, Phys. Rev. Lett. {\bf 57}, 55 (1986); D.G. Caldi
and S. Vafaeisefat, Phys. Lett. B {\bf 356}, 386 (1995).

\bibitem{GSI}
M. Clemente {\it et al.}, Phys. Lett. B {\bf 137}, 41 (1984); T. Cowan
{\it et al.}, Phys. Rev. Lett. {\bf 56}, 444 (1986).

\bibitem{Argonne}
I. Ahmad {\it et al.}, Phys. Rev. Lett. {\bf 75}, 2658 (1995).

\bibitem{Ng}
See, e.g., contribution of Y.J. Ng and Y. Kikuchi in {\it Vacuum
Structure in Intense Fields}, eds. H.M. Fried and B. Muller (Plenum,
New York, 1991); contribution of Y.J. Ng in {\it Tests of Fundamental
Laws in Physics}, eds. O. Fackler and J. Tran Thanh Van (Editions
Frontieres, 33 Gif-sur-Yvette Cedex, 1989), and references therein.

\bibitem{BCT}
A.S. Blaer, N.H. Christ, and J.F. Tang, Phys. Rev. Lett. {\bf 47}, 1364
(1981).

\bibitem{KL}
S.P. Klevansky and R.H. Lemmer, Phys. Rev. D {\bf 39}, 3478 (1989).

\bibitem{GMS}
V.P. Gusynin, V.A. Miransky, and I.A. Shovkovy, Phys. Rev. D {\bf 52},
4747 (1995).

\bibitem{Ritus}
Contribution of V.I. Ritus in {\it Issues in Intense-Field Quantum
Electrodyanamics}, ed. V.L. Ginzburg (Nova Science, Commack, 1987).

\bibitem{math}
See, e.g., {\it Handbook of Mathematical Functions}, eds. M. Abramowitz
and I.A. Stegun (Dover, New York, 1964).

\bibitem{ortho}
The orthonormality property reads  $\int d^4x \bar{E}_{p'}(x) E_p(x) =
(2 \pi)^4 \hat{\delta}^{(4)}(p-p')$ and the completeness property is as
contained in $\Sigma \!\!\!\!\! \int d^4p E_p(x) \bar{E}_p(y) = (2 \pi)^4
\delta^{(4)}(x-y)$.

\bibitem{chi}
In the absence of an electric field, we have $\tilde{E}_{p11} =
\tilde{E}_{p1-1}$ and $\tilde{E}_{p-11} = \tilde{E}_{p-1-1}$
so that references to $\chi$ are irrelevant and can be dropped.

\bibitem{HN}
See, e.g., K. Higashijima and A. Nishimura, Nucl. Phys. {\bf B113},
173 (1976).  Also see Ref. \cite{Mir} and references contained therein.

\bibitem{DW}
Probably chiral symmetry breaking is inhibited by an electric field.
See E. Dagotto and H.W. Wyld, Phys. Lett. B {\bf 205}, 73 (1988);
also see Ref. \cite{Ng}.

\end{thebibliography}
\end{document}